\documentclass[prl,twocolumn,showpacs,preprintnumbers,aps]{revtex4}

\usepackage{graphicx}
\usepackage{bm}

\begin{document}

\preprint{cond-mat/xxxxxx}

\title{Vortex Imaging in the $\pi$-Band of Magnesium Diboride}

\author{M. R. Eskildsen$^{1}$}%
\email{morten.eskildsen@physics.unige.ch}

\author{M. Kugler$^1$}
\author{S. Tanaka$^{1,2}$}
\author{J. Jun$^3$}
\author{S. M. Kazakov$^3$}
\author{J. Karpinski$^3$}
\author{\O . Fischer$^1$}

\affiliation{$^1$DPMC, University of Geneva, 24 Quai E.-Ansermet,
                 CH-1211 Gen\`{e}ve 4, Switzerland\\
             $^2$Department of Physics, Saga University, Saga 840-8502, Japan\\
             $^3$Solid State Physics Laboratory, ETH, CH-8093 Z\"{u}rich,
                 Switzerland}

\date{\today}

\begin{abstract}
We report scanning tunneling spectroscopy imaging of the vortex lattice in
single crystalline MgB$_2$. By tunneling parallel to the $c$-axis, a single
superconducting gap ($\Delta = 2.2$ meV) associated with the $\pi$-band is 
observed. The vortices in the $\pi$-band have a large core size compared to
estimates based on $H_{\text{c2}}$, and show an absence of localized states in
the core. Furthermore, superconductivity between the vortices is rapidly
suppressed by an applied field. These results suggest that superconductivity
in the $\pi$-band is, at least partially, induced by the intrinsically
superconducting $\sigma$-band.
\end{abstract}


\maketitle

Superconductivity in magnesium diboride (MgB$_2$) with a remarkably high
$T_{\text{c}} = 39$ K was recently reported by Nagamatsu {\em et al.}
\cite{nagamatsu}. Since then, great attention has been directed towards
understanding the detailed nature of superconductivity in this material, an in
particular whether this is a one- or two-gap superconductor. Two-gap
superconductivity was predicted theoretically \cite{liu,choi}, and is now
supported by an increasing number of experimental reports
\cite{wang,szabo,chen,giubileo,buzea,bouquet,schmidt,junod,iavarone}. Two-gap
or two-band superconductivity was first studied in the fifties
\cite{gladstone}, and has now found renewed relevance in MgB$_2$. In addition,
and contrary to many materials or alloys studied earlier, the two bands in
MgB$_2$ have roughly equal filling factor, opening the possibility for 
interesting new phenomena. However, the exact microscopic details are still
largely unexplored. An ideal way to address this issue is by local
spectroscopic investigations of the mixed state, which has become possible
with the recent availability of high quality MgB$_2$ single crystals.

In this Letter we report on scanning tunneling spectroscopy (STS) measurements
on single crystal MgB$_2$, including the first vortex imaging in this material.
Tunneling parallel to the $c$-axis, we are able to selectively measure only
the $\pi$-band \cite{brinkman}, in which the vortices are found to have a
number of remarkable properties: An absence of localized states, a very large
vortex core size compared to the estimate based on $H_{\text{c2}}$, and a
strong core overlap.

The STS experiments were performed using a home built scanning tunneling
microscope (STM) installed in a $^3$He, ultra high vacuum cryostat holding a
14 T magnet \cite{kugler}. The measurements were done on the surface of an as
grown single crystal, using electrochemically etched iridium tips. Single
crystals of MgB$_2$ were grown using a high pressure method described elsewhere
\cite{karpinski}, yielding platelike samples with the surface normal parallel
to the crystalline $c$-axis. The surface of the crystals are roughly
$0.25 \times 0.25$ mm$^2$, and the thickness of the order of microns. The
critical temperature is typically $T_{\text{c}} \approx 38 - 39$ K, with a
sharp transition, $\Delta T_{\text{c}} = 0.5$ K, measured by  SQUID
magnetometry \cite{angst}. The STS experiments were done with both the
tunneling direction and the applied magnetic field parallel to the $c$-axis,
and the differential conductivity measured using a standard AC lock-in
technique. In this configuration the upper critical field extrapolates to
$H_{\text{c2}}(T = 0 \text{K}) = 3.1$ T \cite{angst}. Using the Ginzburg-Landau
(GL) expression for $H_{\text{c2}} = \phi_0 / (2 \pi \xi^2)$, where
$\phi_0 = h/2e$ is the flux quantum, yields a coherence length,
$\xi_{\text{GL}} = 10$ nm. An estimate of the mean free path, based on the
measured residual resistivity \cite{sologubenko} and specific heat
\cite{junod}, and the calculated Fermi velocity \cite{brinkman}, gives
$l = 50 - 100$ nm, indicating that the samples are in the clean limit.

We will first focus on the zero-field electronic spectrum of MgB$_2$. The
observation of a single or double gaps depends on the orientation of the
sample, as shown by Iavarone {\em et al.}, who investigated a number of single
grains with different, but unknown absolute orientations \cite{iavarone}. Here
we report the first STS measurements on a MgB$_2$ single crystal, which allow a
correlation between the tunneling direction and the observed gap(s). In Fig. 1
we show a superconducting spectrum obtained at a temperature of 320 mK.
\begin{figure}
\includegraphics*[width=\linewidth]{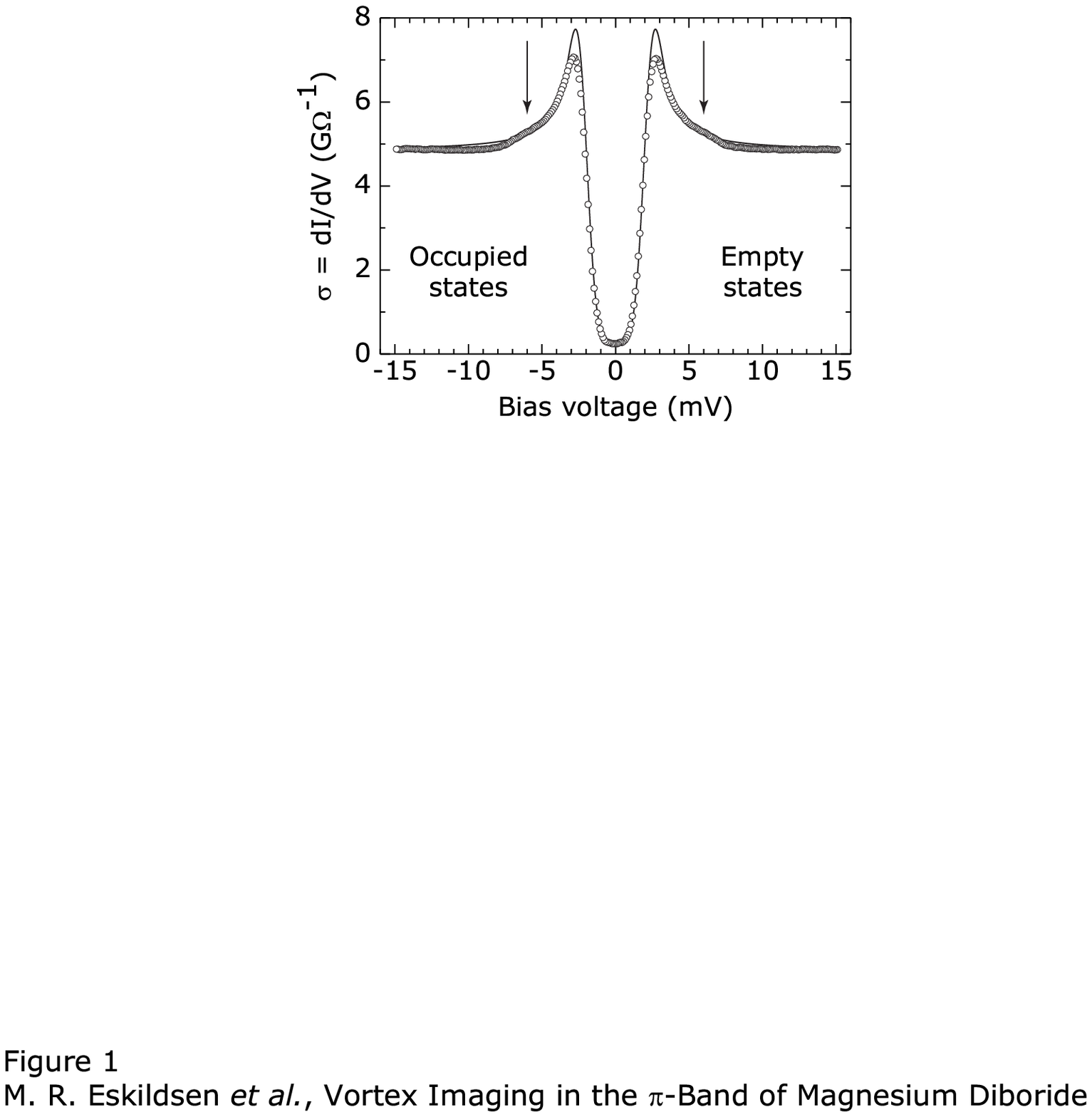}
\caption{
  Zero-field superconducting spectrum of MgB$_2$ at 320 mK for tunneling
  parallel to the $c$-axis and a tunnel resistance, $R_{\text{t}} = 0.2$
  G$\Omega$ ($U = 0.1$ V; $I = 0.5$ nA). The bias voltage is applied to the
  sample. Clear coherence peaks are seen at $\pm 2.5$ meV, and additional weak
  shoulders at $\pm 6$ meV as indicated by the arrows. The line is a fit to the
  Dynes DOS \protect\cite{dynes} ($\Delta = 2.2$ meV, $\Gamma = 0.1$ meV)
  convoluted by a gaussian of width $0.5$ meV RMS to account for experimental
  smearing, including the use of a finite AC excitation ($0.4$ mV RMS).
  \label{fig1}}
\end{figure}
This is an average of 40 spectra obtained along a 100 nm path, which shows
perfect homogeneity. One observes a single gap with coherence peaks at
$\pm 2.5$ meV, and additional weak shoulders at $\pm 6$ meV, as indicated by
the arrows. In addition, the flat region around the Fermi energy proves the
absence of nodes in the gap, and hence that MgB$_2$ is a $s$-wave
superconductor. The very low zero bias conductance indicates a high quality
tunnel junction and a low noise level. True vacuum tunneling conditions were
assured by varying the tunnel resistance, $R_{\text{t}}$, and verifying that
the spectra normalized to the conductance outside the superconducting gap
collapse on a single curve. The spectrum can be fitted by the BCS expression
for the density of states (DOS), including a finite quasiparticle lifetime,
$\Gamma$ \cite{dynes}, and an experimental broadening. The result of the fit is
shown in Fig. 1 and yields a superconducting gap, $\Delta = 2.2$ meV. We have
studied the temperature dependence of the superconducting gap and found
excellent agreement with the BCS $\Delta (T)$.

The fact that only one superconducting gap is observed for tunneling parallel
to the $c$-axis, is explained theoretically by calculations of the Fermi
surface and an analysis of how tunneling along different directions is coupled
to the different bands. The Fermi surface of MgB$_2$ falls into two distinct
sheets: One is derived from $\sigma$-antibonding states of the boron $p_{xy}$
orbitals and is a two-dimensional cylindrical sheet parallel to $c^*$, while
the other consists of $\pi$-bonding and antibonding states of the boron $p_z$
orbitals and is three-dimensional \cite{kortus,liu,choi}. The tunneling matrix
element is different for the two bands, and depends on the tunneling direction.
The $c$-axis tunneling probability into the $\sigma$-band is ten times smaller
than into the $\pi$-band \cite{brinkman}. Furthermore, the calculated
superconducting gap sizes for the two different Fermi surfaces are different,
with $\Delta_{\sigma} \approx 7$ meV, and $\Delta_{\pi} \approx 2$ meV
\cite{choi}. This is in agreement with our results, where one gap with
$\Delta = \Delta_{\pi} = 2.2$ meV is observed, with the shoulders at 6 meV
being a remnant of $\Delta_{\sigma}$. The selective sensitivity to
$\Delta_{\pi}$ turns out to be particularly useful, as we will show in the
following.

We now turn to measurements in an applied magnetic field. In a type-II
superconductor such as MgB$_2$, a magnetic field penetrates into the sample
in the form of vortices each carrying one flux quantum, which are generally
arranged in a periodic array: the vortex lattice. In the core of each vortex,
superconductivity is suppressed within a radius roughly given by the coherence
length, $\xi$. The vortex spacing is determined by the applied field and the
flux quantization, and in the case of a hexagonal vortex lattice it is
$d = (2/\surd 3 \; \phi_0/H)^{1/2}$. The magnetic fields were applied at 2 K,
and the system allowed to stabilize for at least a few hours. After this time
no vortex motion was observed, indicating a fast relaxation and hence low
vortex pinning in the crystal.

In Fig. 2a we show a STS image of a single vortex induced by a field of
$0.05$ T.
\begin{figure}
\includegraphics*[width=\linewidth]{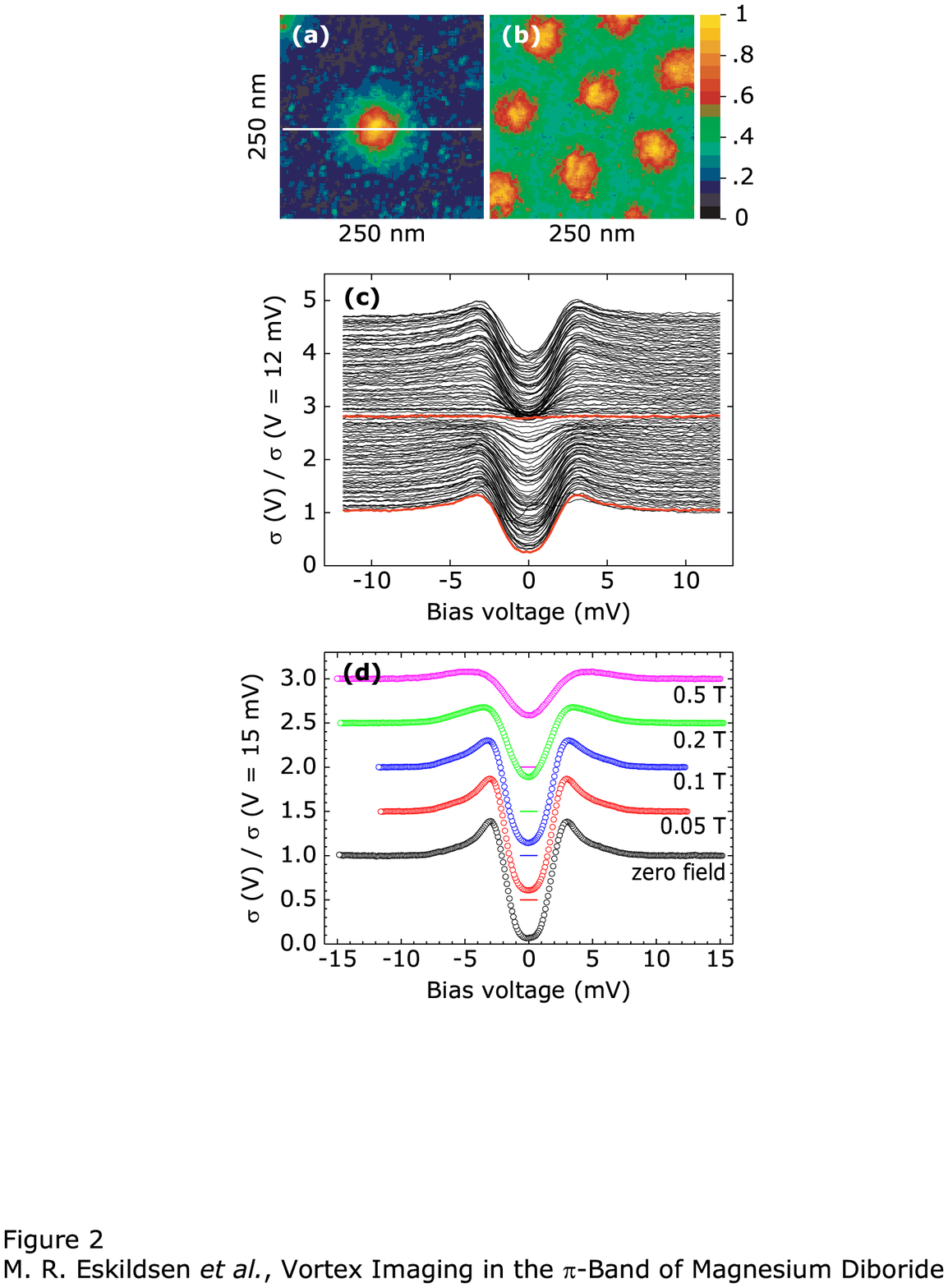}
\caption{
  (color) Vortices in MgB$_2$.
  Top: $250 \times 250$ nm$^2$ false color spectroscopic images of a single
  vortex induced by an applied field of $0.05$ T (a), and the vortex lattice
  at $0.2$ T (b). In both cases the tunnel resistance was,
  $R_{\text{t}} = 0.4$ G$\Omega$ ($U = 0.2$ V; $I = 0.5$ nA). The conductance
  is normalized to respectively $2.9$ meV ($0.05$ T) and $3.9$ meV ($0.2$ T).
  (c): 250 nm trace across the single vortex indicated by the white line in
  (a), with spectra recorded each 2 nm. Each spectrum is normalized to the
  conductivity at 12 meV, and $R_{\text{t}} = 0.4$ G$\Omega$. A spectrum at the
  vortex centre together with one far from the vortex core have been
  highlighted in red for clarity.
  (d): Normalized spectra measured in zero field, and between the vortices for
  fields between $0.05$ T and $0.5$ T ($R_{\text{t}} = 0.2$ G$\Omega$). Each
  subsequent spectrum is offset by $0.5$ with respect to the previous one. The
  bars at zero bias indicate the respective zero conductivity for the offset
  spectra.
  All measurements in this figure were performed at 2 K.
  \label{fig2}}
\end{figure}
The image was obtained by measuring the differential conductance at
zero bias and normalizing this to the conductance at the coherence peak. Low
values of the normalized conductance correpond to superconducting areas, and
high values to the vortex cores. The low field is equivalent to a separation,
$d = 220 \text{ nm} \gg \xi$. The vortices can therefore be considered as
isolated from each other. Such isolated vortices are expected to contain
localized quasiparticle states, which should show up as a zero bias conductance
(ZBC) peak at the vortex centre \cite{hessgygi}, provided that the sample is
sufficiently clean to prevent these to be smeared out by scattering. We have
measured the evolution of the spectra at a large number of positions along a
trace across the vortex core as shown in Fig. 2c. Contrary to expectations, we
find that the normalized ZBC increases to one with no indication of any
localized states. Instead, the spectra in the centre of the vortex are
{\em absolutely flat}, with no excess spectral weight at or close to zero bias.
This absence of localized states is striking, considering that
$l = 5 - 10 \times \xi_{\text{GL}}$. However, as we will show below, the
coherence length in the $\pi$-band is approximately 50 nm. This is much larger
that the estimate based on $H_{\text{c2}}$, and equal to only one to two times
the mean free path. Nonetheless, systematic studies of Nb$_{1-x}$Ta$_x$Se$_2$
with $x = 0 - 0.2$, showed that even for $\xi/l \approx 1$ some excess weight
close to zero bias was observed \cite{renner}. In parallel to STS imaging, STM
topographic images were recorded (not shown), which revealed a flat surface
with a RMS roughness of 6 \AA \ over the whole image area.

Before analyzing the vortex profile in detail, we will consider the situation
at higher fields. In Fig. 2b we show the STS image of the hexagonal vortex
lattice observed at $0.2$ T. We notice that the normalized ZBC between the
vortices, is now enhanced with respect to the value far from the single vortex
at $0.05$ T. This increase of the ``bulk'' ZBC is unusual at such a modest
field, only about 7\% of $H_{\text{c2}}$. To elucidate this behaviour, ``bulk''
spectra for fields between zero and $0.5$ T are shown in Fig. 2d. It is clear
that even modest fields rapidly suppress superconductivity in the region
between the vortices. This is seen both by an increase of the ZBC and by a
suppression of the coherence peaks outside the vortex cores, which one would
only expect for fields close to $H_{\text{c2}}$, corresponding to a significant
core overlap. This is consistent with earlier point contact spectroscopy
measurements \cite{szabo}, with the addition that we resolve the local
behaviour on a microscopic scale.

We now return to the single vortex measurement. In Fig. 3a, we have plotted
the normalized ZBC, $\sigma'(x,0)$, for the vortex trace measured at $0.05$ T.
\begin{figure}
\includegraphics*[width=\linewidth]{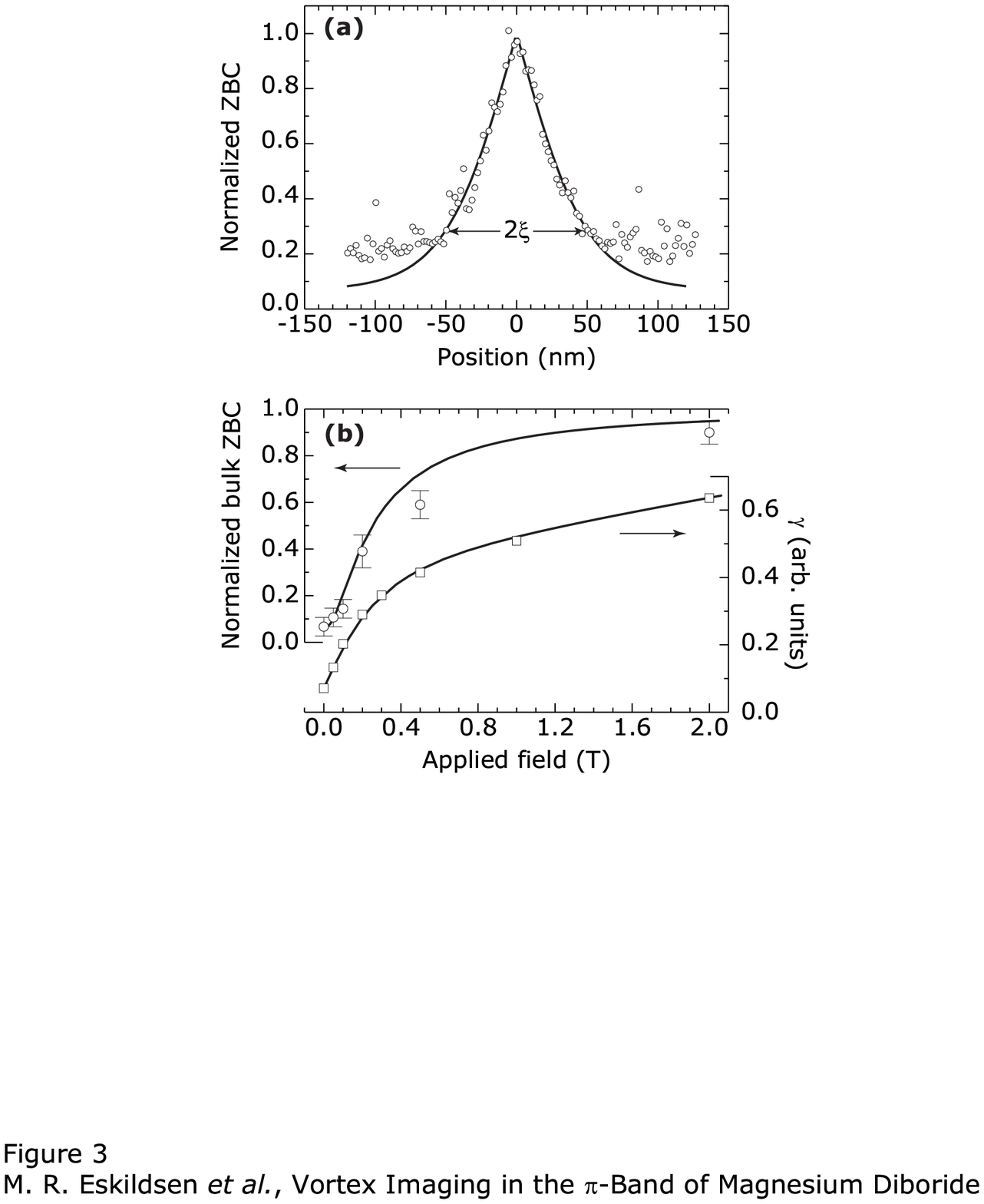}
\caption{
  Vortex size and core overlap.
  (a): Normalized zero bias conductance versus distance from the centre, for
  the isolated vortex shown in Fig. 2a. The line is a fit to eq. (\ref{sigma})
  in the text.
  (b): Calculated ``bulk'' ZBC (left axis) and electronic specific heat,
  $\gamma$, for
  $\gamma^{\pi}_{\text{n}}/\gamma^{\sigma}_{\text{n}} = 0.55/0.45$ (right
  axis). The calculated values are compared to respectively the measured bulk
  ZBC (circles), and specific heat measurements on polycrystalline samples
  (squares) \protect\cite{junod}.
  \label{fig3}}
\end{figure}
It is immediately clear that the spatial extension of the vortex core is much
larger than the 10 nm estimated from $H_{\text{c2}}$. Somewhat surprisingly, we
find that the ZBC profile can be fitted by one minus the GL expression for the
superconducting order parameter:
\begin{equation}
  \sigma'(x,0) = \sigma'_0 + (1 - \sigma'_0) \times (1 - \tanh x/\xi),
  \label{sigma}
\end{equation}
where $\sigma'_0 = 0.068$ is the normalized ZBC measured in zero field. The
fit, shown in Fig. 3a, yields a coherence length,
$\xi = \xi_{\pi} = 49.6 \pm 0.9$ nm. Using the GL expression to calculate the
upper critical field with this value of the coherence length, yields
$H'_{\text{c2}} = 0.13$ T. At $0.2$ T we hence find ourselves in the bizarre
situation of imaging the vortex lattice above the nominal $H'_{\text{c2}}$.
Additional vortex lattice imaging has been performed as high as $0.5$ T.

This apparent paradox can be reconciled if one assumes that superconductivity
in the $\pi$-band is induced by the $\sigma$-band, either by interband
scattering, or Cooper pair tunneling \cite{gladstone,nakai}. This means that
isolated the $\pi$-band would either be non-superconducting or have a very low
upper critical field. Consequently the observed behaviour reflects the state
in the $\sigma$-band by an interband proximity effect, along the lines of
recent theoretical work \cite{nakai}.
The vanishing of $\Delta_{\pi}(T)$ at the bulk $T_{\text{c}}$ further supports
this conclusion. Finally, it is also consistent with estimates of the coherence
lengths, using the BCS expression
$\xi_0 = \hbar v_{\text{F}} / (\pi \Delta(0))$ and considering each band
separately. Taking the calculated average Fermi velocity in the $ab$-plane for
the $\pi$-band, $v_{\text{F}}^{\pi} = 5.35 \times 10^5$ m/s \cite{brinkman},
and the measured gap value $\Delta_{\pi} = 2.2$ meV we get $\xi_0^{\pi} = 51$
nm, in excellent agreement with $\xi_{\pi}$ obtained from the vortex profile.
A similar analysis for the $\sigma$-band, using
$v_{\text{F}}^{\sigma} = 4.4 \times 10^5$ m/s \cite{brinkman} and
$\Delta_{\sigma} = 7.1$ meV \cite{iavarone} yields $\xi_0^{\sigma} = 13$ nm.
This agrees with the coherence length obtained from $H_{\text{c2}}$, and
reinforces the conclusion that it is mainly the $\sigma$-band which is
responsible for superconductivity in MgB$_2$, and thus determins the
macroscopic parameters $T_{\text{c}}$ and $H_{\text{c2}}$.

As described above, it is the transfer from the $\sigma$-band that makes
superconductivity in the $\pi$-band possible, despite a strong vortex core
overlap already at very low magnetic fields. Constructing a simple model for
the core overlap in the $\pi$-band by
\begin{equation}
  \sigma'(\bm{r},0) =
    \sigma'_0 + 
    (1 - \sigma'_0) \times 
           \left( 1 - \Pi_i \tanh \frac{|\bm{r} - \bm{r}_i|}{\xi_{\pi}}
           \right),
  \label{overlap}
\end{equation}
where $\bm{r}_i$ are the vortex positions for a hexagonal lattice with a
density corresponding to the applied field, we can calculate the ZBC at any
position in the vortex lattice unit cell. In Fig. 3b we compare the calculated
conductivity at the midpoint between three vortices with the measured bulk ZBC.
This shows a very good agreement, especially taking into account that there are
no free parameters in the calculation: $\xi_{\pi}$ is determined from the
vortex profile, and $\sigma'_0$ from the zero field measurement. The vortex
core overlap also explains the, strongly nonlinear field dependence of the
electronic specific heat, $\gamma$ \cite{wang,junod}. In strongly type-II
superconductors core overlap is usually negligible, with each vortex creating
the same number of quasiparticles at the Fermi surface, and hence contributing
by the same amount to the specific heat. In that case
$\gamma = \gamma_{\text{n}} \; H/H_{\text{c2}}$, where $\gamma_{\text{n}}$
is the electronic specific heat in the normal state. However, in the case of
MgB$_2$, with strong core overlap in the $\pi$-band, the isolated vortex
assumption is violated. Instead, one can calculate the contribution from the
$\pi$-band, simply by averaging the normalized ZBC in one vortex lattice unit
cell,
$\gamma_{\pi} = \gamma^{\pi}_{\text{n}} \; \langle \sigma'(\bm{r},0) \rangle$
\cite{nakai}. On the other hand, the $\sigma$-band can be described by the
usual linear field dependence
$\gamma_{\sigma} = \gamma^{\sigma}_{\text{n}} \; H/H_{\text{c2}}$. Adding the
terms gives $\gamma = \gamma_{\pi} + \gamma_{\sigma}$, where
$\gamma^{\pi}_{\text{n}}/\gamma^{\sigma}_{\text{n}}$ is the relative weigth of
the two bands. The calculated field dependence of $\gamma$ is shown in Fig. 3b,
in perfect agreement with the measured specific heat for polycrystalline
MgB$_2$ \cite{junod}, using
$\gamma^{\pi}_{\text{n}}/\gamma^{\sigma}_{\text{n}} = 0.55/0.45$.

In summary, we have presented STS data on the $\pi$-band in MgB$_2$, including
the first vortex imaging in this material. We have demonstrated the absence of
localized states in the vortex core, a very large vortex core size and a strong
core overlap. The data presents a striking experimental demonstration of the
fundamentally different microscopic properties of the two bands in MgB$_2$.

We acknowledge valuable discussions and communication of data prior to
publication with F. Bouquet, Y. Wang and A. Junod, and thank B. W. Hoogenboom
and I. Maggio-Aprile for sharing their experience in STM/STS. This work was
supported by Swiss National Science Foundation. M.R.E. has received support
from the Christian and Anny Wendelbo foundation and from The Danish Natural
Science Research Council.


\end{document}